# X-Ray Spectral Parameters for a Sample of 95 Active Galactic Nuclei

A.A. Vasylenko[1], V.I. Zhdanov[2], E.V. Fedorova[2]
[1]Main Astronomical Observatory NAS of Ukraine, Kyiv, Ukraine
e-mail: merak@ukr.net
[2]Taras Shevchenko National University of Kyiv, Ukraine

*We present a broadband X-ray analysis of a new homogeneous sample of 95 active galactic nuclei (AGN) from the 22-month Swift/BAT all-sky survey. For this sample we treated jointly the X-ray spectra observed by XMM-Newton and INTEGRAL missions for the total spectral range of 0.5-250 keV. Photon index $\Gamma$, relative reflection R, equivalent width of Fe $K_\alpha$ line $EW_{FeK}$, hydrogen column density $N_H$, exponential cut-off energy $E_c$ and intrinsic luminosity $L_{corr}$ are determined for all objects of the sample. We investigated correlations $\Gamma$–R, $EW_{FeK}$– $L_{corr}$, $\Gamma$–$E_c$, $EW_{FeK}$–$N_H$. Dependence "$\Gamma$ – R" for Seyfert ½ galaxies has been investigated separately. We found that the relative reflection parameter at low power-law indexes for Seyfert 2 galaxies is systematically higher than for Seyfert 1 ones. This can be related to an increasing contribution of the reflected radiation from the gas-dust torus. Our data show that there exists some anticorrelation between $EW_{FeK}$ and $L_{corr}$, but it is not strong. We have not found statistically significant deviations from the AGN Unified Model.*

## 1. INTRODUCTION

It is widely accepted today that the variety of observational appearances of AGNs can be divided in two big groups: those caused by physical conditions in particular AGN, and those due to geometrical effects of the AGN orientation with respect to the line-of-sight. The latter effects are well explained by the famous Unification Model (UM, Antonucci 1993). According to the UM, we can observe the variety of Seyfert types of AGNs due to a presence of toroidal absorbing structures surrounding the AGN central parts. Orientation of such structure with respect to the line-of-sight causes observational differences between AGNs of different Seyfert types. This model is consistent with numerous X-ray observations, confirming that S2 AGNs have a higher absorption values than in S1 ones (see, e.g., Awaki 1991; Dadina 2008; Singh et al. 2011).

However, more than 10 years ago, some deviations from UM have been found as well. Firstly, some Seyfert 2 types have too low level absorption, comparable with the absorption level typical rather to Sy1 type AGNs (Dadina 2008; Panessa & Bassani 2002). On the other hand, some Sy1 AGNs, have a comparatively high level of absorption (Cappi et al. 2006). Secondly, photon index is skewed towards low values for AGNs with significant absorption (Panessa et al. 2008; Vasudevan et al. 2013). Also, as it was repeatedly claimed that cut-off energy for Seyfert 2 type galaxies is 1.5-2 times more as compared to type 1 (Deluit & Courvoisier 2003; Dadina 2008; Beckmann et al. 2009). These observational peculiarities cannot be explained by UM, and this requires considering specific features of processes in the AGN nucleus.

As widely accepted, central sources of radiation in AGNs are accretion disks and their coronae near super-massive black holes, and the X-ray emission is primarily produced via unsaturated inverse Compton scattering of thermal ultraviolet photons of the accretion disc by hot (may be, thermal) electrons of corona (see e.g., Haardt & Maraschi 1991, 1993; Done 2010). This radiation can have different spectral properties depending on the geometry of the corona, its condition and state of the accretion (Beloborodov 1999; Perola et al. 2002; Petrucci et al. 2001; Ricci et al. 2011; Vol'vach et al. 2011). The presence of the gas-dust torus around the nucleus also strongly affects the absorption and emission lines.

The main spectral features in the X-rays are the photon index, the cut-off energy, the relative reflection, and the most important and brightest neutral Fe $K_\alpha$ emission line (at 6.4 keV in the local reference frame). The last three observational parameters depend on the state of the accretion disk and of the presence of a gas-dusty torus on the line of sight (Awaki 1991; Cappi et al. 2006; Petrucci et al.

2001; Ueda et al. 2007). The torus optical thickness and its geometry shape reveal themselves in the anisotropy of the absorption and scattering (Ricci et al. 2011; Ueda et al. 2007). Note that the photoelectric cross-section sharply declines above 20-30 keV, and have a significant influence only in soft and middle energy <2-3 keV.

Therefore, the hard X-rays are most relevant for investigating the intrinsic emission from core of AGN and to test UM, in the sense that this energy range directly provides information about the state of the accretion disk and it corona. Also, the important feature is the reflection hump with a peak at 20-30 keV energy that may represent not only physical state of the accretion disk (Magdziarz & Zdziarski 1995), but also some information about the gas-dust torus form. At low energies (up to ~3 keV) spectrum contains information about the features of absorption medium, both cold (that is related to the gas-dust torus), and warm (that are related with either broad or narrow line region). In addition, Fe $K_\alpha$ 6.4 keV emission line parameters depend on of the characteristics both of the accretion disk and gas-dust-torus. Thus, investigation of the inter-dependencies and cross-correlations of X-ray spectral parameters of continuum and such features as the absorption (column densities), the equivalent widths of emission lines, relative reflection, intrinsic luminosity and the photon index will allow us to analyse a number of the physical processes in AGN, which cannot be observed directly.

The most prominent physical difference between AGNs of various classes may be related with jet activity and RL/RQ dichotomy. The model to explain such physical differences in AGN structure is known as a "spin-paradigm" because the occurrence of jets is associated with high spin values of the central black hole in AGN and with the direction of rotation of the accretion disk around the black hole (Garofalo et al. 2010). This model predicts the low values of the exponential cutoff of the primary emission of an RL AGN (below 100 keV) and high values or absence of it in an RQ AGN spectrum. However, there are some objects that seem to contradict this paradigm (De Rosa et al. 2008, Fedorova et al. 2011, Soldi et al. 2005, 2011). That is why it is interesting to analyse the dependence of the high-energy exponential cut-off from the RL/RQ characteristics of the AGNs.

In the present paper we derive the main parameters of the broad-band X-ray spectra (0.5-250 keV) of 95 Seyfert galaxies on the basis of an original sample available from both INTERGAL/IBIS and XMM-Newton/EPIC observations. Because of a comparatively large number of data we were able to perform a statistical treatment that involves correlations of the spectral parameters, and analyse how they depend on the Seyfert type and RL/RQ class with the final aim to check how they match the UM predictions.

In Sections 2 and 3 we describe the homogeneous sample of AGNs and the data analysis that has been performed. Note that most of the sources treated in the sample have been studied elsewhere, including with the aim of testing UM (see, e.g., De Rosa et al. 2008; Molina et al. 2009; Chesnok et al. 2009; Ricci et al, 2011; De Rosa et al. 2012; Vasudevan et al. 2013; Pulatova et al. 2015). In order to specify main differences of our paper from the earlier studies, we note the following. De Rosa et al. (2012) have studied in detail the AGN with large absorption; the dependencies of $N_H$ upon different parameters have been investigated as well as the behavior of luminosities for 2-10 and 20-100 keV energy ranges. Ricci et al (2011) deal essentially with average hard X-ray spectra based only on the INTEGRAL data. Molina et al. 2009 performed a detailed analysis with larger AGN sample, though for Seyfert1s only. Recent paper by Molina et al. 2013 deals with even larger sample based on the INTEGRAL/IBIS and Swift/BAT observations. Vasudevan et al. (2013) use for low-energy (<10 keV) spectral analysis the XMM-Newton+Swift/XRT data and for high-energy (14-195 keV) – Swift/BAT data. However, most of the earlier studies do not cover *simultaneously* all 0.5-250 keV range that can affect the interpretation of observed characteristics of the Fe K line (6.4 keV), and correlations with the characteristics of the hard X-ray part of the spectrum. De Rosa et al. 2008, 2012 deal with the same energy range as we do (see below), however for much smaller samples (7 and 33 objects). Molina et al. (2009, 2013) consider 1-110 keV and 17-150 keV with aims different from ours.

In Sect. 3, we present analysis of relations between different X-ray spectral parameters; a statistical analysis of the parameters and their comparisons for two Seyfert types are discussed in Sect.

4 (Tables 1 and 2). The individual parameters of the objects derived in our treatment are listed in the Tables 3 and 4.

## 2. DATA REDUCTION

We compiled a sample of galaxies with active nuclei of the 22-months Swift/BAT All Sky Survey for energies 15-195 keV, namely, we used the catalog (Tueller, J. et al. 2010). We dealt with the objects that are already identified as AGNs according this catalog and to NED (https://ned.ipac.caltech.edu/, and we checked the literary sources of this database in case of any ambiguity). All double X-ray systems, blazars and unidentified objects have been removed. Using the catalogue data allowed us to avoid in part several factors such as influence of star formation (in case of optical or UV catalogues) and effects due to limited sensitivity of BAT detectors at low energies. These factors can cause bias of the sample for a certain class of AGNs. We believe that our cleaned sample is well suited for analysis of various Seyfert types, AGN X-ray radiation characteristics, and, therefore for testing of UM. We selected only those galaxies for which we were able to build both X-ray spectra observed with both XMM-Newton and INTEGRAL satellites. This allowed us to work within 0.5-250 keV energy range. The final sample contains 95 galaxies, including 54 Seyfert1s (1-1.5) and 41 Seyfert2s (1.8-2). The sample includes 25 radio-loud galaxies and 70 radio-quiet ones identified according to NED and literary sources of this database.

Here we used the traditional definition of radio loudness using boundaries for the radio flux at 5 GHz ($P_{5\ GHz}$, Miller, Peacock & Mead 1990) and the ratio between radio at 1.4 GHz and optical B-band flux density (Kellerman et al. 1989). According to this criterion, a galaxy is radio quite if $P_{5\ GHz} \leq 10^{32}$ erg s$^{-1}$ Hz$^{-1}$ (or $P_{5\ GHz} \sim 10^{25}$ W Hz$^{-1}$ sr$^{-1}$) and $F_{5\mathrm{GHz}} / F_{B-band} < 10$, respectively. For several objects (in the absence of data for radio-optical criterion) we also used the X-ray criterion, as ratio between 2-10 keV and 20-100 keV intrinsic luminosities and radio luminosity at the above frequencies (see, e.g., Terashima & Wilson 2003; Panessa et al. 2007) and the IR criteria at 25.89-12.81 micrometers (see, e.g., Melendez, Kraemer & Schmitt 2010; Lacy, Ridgway & Gates 2013). To be exact, all radio loud AGNs of our sample satisfy criterion $F_{5\mathrm{GHz}} / F_{B-band} > 10$, except several objects listed as follows. These are: FR II galaxy 4C 50.55 where we rely upon the detailed study by Molina et al. (2005); in case of 3C 452 we used the $IR_{15\mu}$ criterion and in case of NGC 788 and WKK 6471 we used the X-ray criterion. There is the only Seyfert 1.9 galaxy NGC 5252 in our sample, which has an intermediate values of X-ray ($L_{5\mathrm{GHz}} \big/ L^{\mathrm{int}}_{2-10\,\mathrm{keV}} = -4.7$) and radio ($F_{5\mathrm{GHz}} \big/ F_{B-band} \approx 2$) criteria, but it was reported that this galaxy has radio jets (Liu & Zhang 2002).

The X-ray data for the sample were processed using the standard software packages XMM SAS ver. 11.0 (Science Analysis Software) according to the guidelines of XMM-Newton User's Manual. Given its higher sensitivity, we use preferably the time averaged EPIC/PN spectrum for the analysis of each object, except for observations, where duration of exposure for EPIC/MOS has been significantly longer (e.g. for NGC 4074). Only patterns corresponding to single and double events (pattern≤4) were taken into account for the PN camera (pattern≤12 for MOS1/2 cameras). Filter FLAG=0 was applied for excluded bad pixels and events that are at the edge of a CCD. The ARFGEN and RMFGEN tasks were used to create ancillary and response files. The source spectra were extracted from a circular region typically 30-40 arcsec centered on the source, while the background regions have been chosen to be free from contaminating sources. Exposures for all objects have been filtered for periods of high background using tabgtigen task. Spectra were binned according to the luminosity of each source.

In order to obtain the hard X-ray spectra we treated 50 longest SCW (Science window) INTERGAL/IBIS data for every object. INTEGRAL data analysis has been performed using standard procedures of OSA 9.0 (Offline Standard Analysis Software), the ibis_science_analysis metatask. All the spectra were extracted individually for every science window and then summed up into the whole set, using the spe_pick metatask.

The XMM-Newton and INTEGRAL/IBIS spectra were treated together and analyzed using XSpec ver.12.6 software. Since this observations are not simultaneous, a cross-calibration constants C have been introduced in our models.

## 3. X-RAY SPECTRAL ANALYSIS

For the primary spectrum of corona we used the standard power law with an exponential cut-off at high energies $A(E) = KE^{-\Gamma}\exp(-E/E_{cut\text{-}off})$, $E_{cut\text{-}off}$ stands for the cut-off energy. The radiation with this spectrum is reflected from a neutral material of an accretion disc or gas-dust torus; it is subject to further absorption and superimposed with the initial primary spectrum. To account for presence of reflection, we used also pexrav model (Magdziarz & Zdziarski 1995). The following models were involved during the fitting procedure: pexrav (reflection plus power-law continuum); zphabs or absori/zxipcf (with CvrFact=1), that takes into account the neutral or ionized absorption, and its analogue zpcfabs (zxipcf) taking into account the covering factor for neutral or ionized absorption medium respectively; zgauss (gaussian profile of emission lines). Galactic absorption (i.e. in Milky Way) is included in all these spectral models. Thus the model in XSPEC schematically looks like:

$$\mathbf{phabs(N_{HGal})*\{\ Soft\ component+zphabs*[pexrav(\Gamma,\ E_{cut\text{-}off},\ R) + zgauss]\}.}$$

Additional components of neutral or ionized absorption were added into soft/hard component during a fitting in the each individual case. Note that fitting with absori and zxipcf yields practically the same results in models zphabs та zpcfabs for all objects, except for two objects: NGC7469 and LEDA 75476 (without essential effect for the statistical results). In order to derive the luminosities we used the standard ΛCDM cosmological model with parameters $H_0 = 70\,km \cdot s^{-1}Mpc^{-1}$, $\Lambda = 0.73$, $\Omega_m = 0.27$ (Bennett 2003).

For each galaxy of the sample, we built the X-ray spectra and obtained the spectral parameters such as power-law index $\Gamma$, relative reflection parameter $R$, equivalent width of Fe $K_\alpha$ line ($EW_{FeK}$), intrinsic luminosity $L_{corr}$, cut-off energy $E_{cut-off}$ and intrinsic absorption value. The best-fit parameters are presented in Tables 3 and 4. The errors, lower and upper limits quoted correspond to 90% confidence level for the interesting parameter ($\Delta\chi^2 = 2.71$). The best-fitting parameters for the whole sample are reported in Table 3,4. Below we present correlations of the main spectral parameters obtained on this basis, such as "photon index - relative reflection parameter", "equivalent line width of Fe $K_\alpha$ line - intrinsic luminosity" (Baldvin effect), "photon index - cut-off energy" and "absorption value - equivalent line width of Fe $K_\alpha$ ".

## 4. MAIN RELATIONS OF X-RAY PARAMETERS

### 4.1 Photon index - relative reflection

For the first time the relationship between relative reflection R and photon index was found and described by Zdziarski et al. (1999). This work used the X-ray Ginga satellite data for 23 radio-quiet Seyfert galaxies of type 1, 1.2, 1.5 and several X-ray binary systems. The energy range was 1,7-20 keV. They found the value of Spearman correlation test to be 0.91, which is rather a high value. In recent works (De Rosa et al. 2012, Molina et al. 2009, Panessa et al. 2008, Vasudevan et al. 2013) with better quality data this result has not been confirmed. In this view, we deal with a subsample of 58 galaxies (of all Seyfert types) for which we were able to obtain both power-law index and relative reflection parameters. To look for possible specific features appropriate for galaxies of different types, we divided the sample into Seyfert2s (Seyfert 2 and 1.9) Seyfert1s (including Seyfert 1, 1.2, 1.5), radio-quiet and radio-loud.

At the first step we obtained the Spearman correlation coefficient ρ, which, as it turned out, gives a non-zero value[1] $\rho = 0.36 \pm 0.11$ for all 78 active galactic nuclei. For Seyfert1s only, the correlation somewhat increases to $r = 0.47 \pm 0.13$, and for Seyfert2s it is $\rho = 0.22 \pm 0.20$. Indeed, as it is shown in Figs. 1 and 2, for values of $\Gamma > 2$ there is some increase of $R$, that may be viewed as a hint for a relation between $R$ and $\Gamma$). As we see in Fig. 1, for photoindex $\Gamma < 1.5$ the reflection *R* for Seyfert2s tends to be larger as compared to Seyfert1s. Fig. 2 shows that for radio-loud galaxies, the reflection *R* tends to be smaller as compared to radio-quiet ones (cf. Molina et al. 2008).

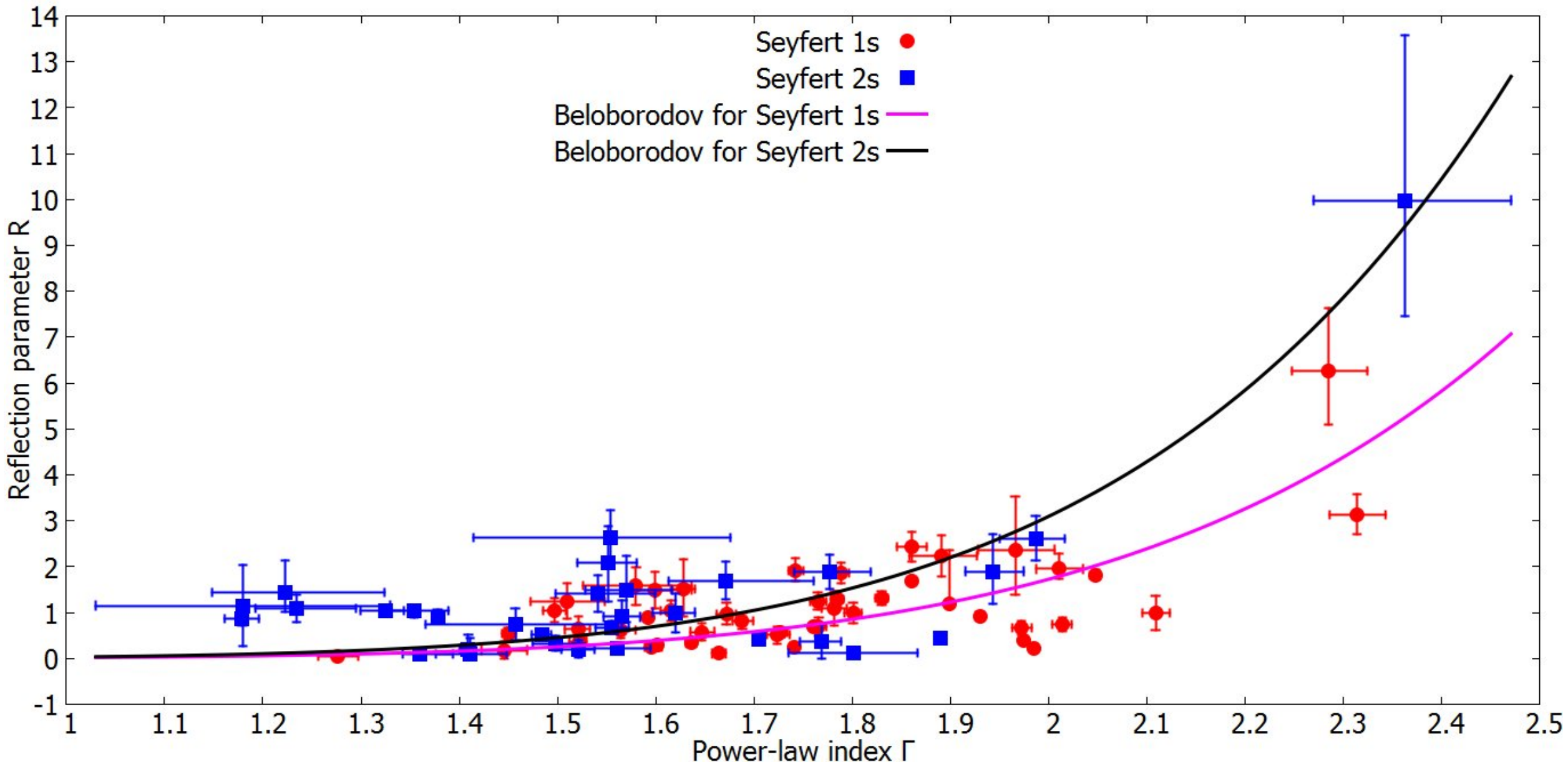

**Fig. 1.** Photon index Γ vs relative reflection coefficient *R* separately for different Seyfert types. Solid lines represent the Beloborodov (1999) model.

The correlation coefficients and the probability of "null hypothesis" (that the parameters are completely independent) are shown in Table 2. Therefore, as a possible option, we examined a fit of this dependence with the Beloborodov model (Beloborodov 1999). In this model, the decrease of X-ray reflection from the disc occurs due to the bulk motion of the X-ray-emitting hot plasma in the direction of the reflector. The behaviour of the data for the small values of the photon index is rather stochastic; whereas, the amount of the data is insufficient for confident conclusion that this dependence exists for high $\Gamma$.

In summary, it can be argued that our sample data does not allow us to present a definite answer to the question about the relationship between the photon index and the relative reflection. For identifying a possible dependency, at least, (i) we need a large number of high-quality X-ray data on galaxies, (ii) we must use a realistic model of the accretion disk coronae and (iii) we must separate the reflection of the gas-dust torus and that of the accretion disk.

---

[1] The errors of ρ are jackknifed; here and below they correspond to 1σ limits.

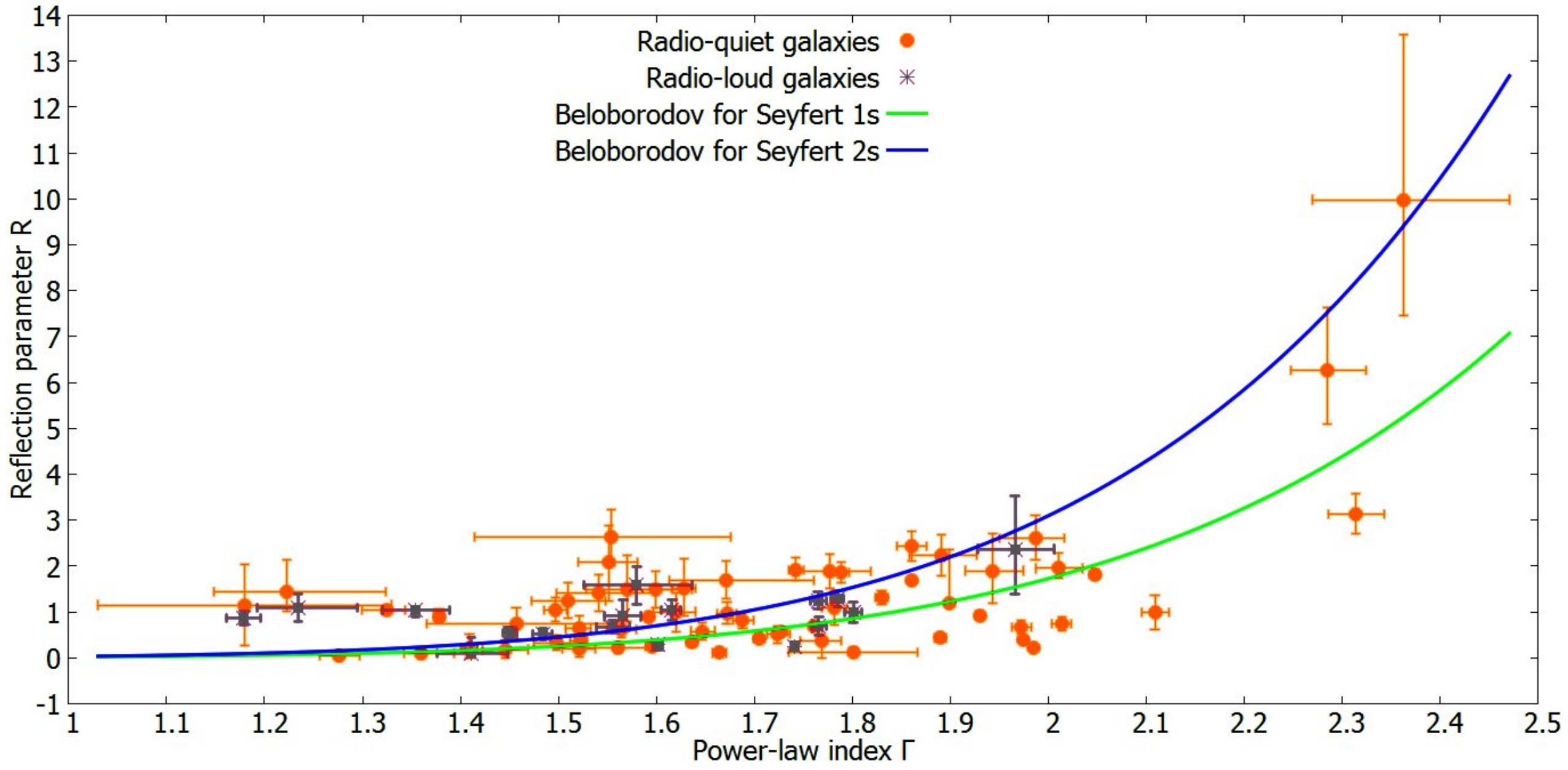

**Fig. 2.** Photon index Γ vs relative reflection coefficient *R* separately for different radio types. Solid lines represent the Beloborodov (1999) model.

### *4.2 Absorption value - equivalent line width of Fe K$_\alpha$*

The dependence of Fe K$_\alpha$ equivalent line width upon $N_H$ is often used to identify where the Fe Kα line is emitted: either in the accretion disk or in the gas-dust torus (e.g. De Rosa et al 2012, Page et al. 2004). To estimate this dependence we used the data for 68 galaxies, mostly of the Seyfert 2 type.

Up to the $N_H = 10^{23.5} cm^{-2}$ the equivalent width does not change significantly (see Fig. 3). Its average value is $EW_0 = 92.5 \pm 9.8$ eV. Above $N_H = 10^{23.5} cm^{-2}$ the situation is different, showing an increase of the equivalent width as $N_H$ grows. Awaki et al. (1991) described this behaviour theoretically taking into account the reflection of X-rays from the Compton-thick torus. For example, the large equivalent width with $N_H \sim 10^{23.5} \div 10^{24} cm^{-2}$ could originate in a thick torus (which is also responsible for the observed absorption). Ghisellini et al. (1994) showed that a typical gas-dust torus with $N_H = 10^{24} cm^{-2}$ should produce iron $EW \sim 650$ eV for the solar abundance. At the same time, the large equivalent width at low $N_H$ $(< 10^{23.5} cm^{-2})$ suggests that the Fe K$_\alpha$ line is formed due to a reflection from a material different from the absorber. Small equivalent width (~ 100 eV) for galaxies with low absorption and Seyfert1s mean that the lines (not all, but with large $\sigma_{line} \geq 0.3$ keV) are formed in the same environment. Note that the broad line region is also discussed as a possible source of the Fe K$_\alpha$ line (cf. De Rosa et al. 2012, Page et al. 2004, Nandra 2006). For the absorption less than $10^{23} cm^{-2}$ the value $EW_{FeK} > 100$ eV cannot be explained by a reflection from the gas-dust torus, which is responsible for the neutral absorption (cf., e.g., Yaqoob et al. 2001). There must be an additional region (besides the accretion disk) – this is BLR (possibly, some part of it, which is closer to the black hole; though some authors state that one must consider overall BLR). Also, Ricci et al (2014a) present arguments that the bulk of the narrow Fe Kα line is produced by the same material as that responsible for the mid-infrared emission.

In our case the solid line in Fig. 3 represents the function $EW(N_H) = EW_0 \exp(\sigma_{Fe} N_H)$ and well reproduces the behaviour of the equivalent width, which increases with the absorption. This curve represents the case when the Fe K$_\alpha$ line is formed by a reflection from the broad line region in the presence of the absorber, which *does not lie exactly on the lines of sight* of galaxies with low $N_H$, however it absorbs only a part of the continuum emission. If the latter is really the case, this allows us to explain the value of $EW_{FeK} \sim 50 \div 150$ eV with $N_H < 10^{23.5} cm^{-2}$.

To summarize, we note that our results are consistent with the concept that in the galaxies with the high absorption ($N_H > 10^{23.5}$ $cm^{-2}$), the Fe $K_\alpha$ lines can be generated by gas-dust torus and for the low values of absorption ($N_H < 10^{23.5}$ $cm^{-2}$) the Fe $K_\alpha$ line is mostly generated in an environment close to the black hole.

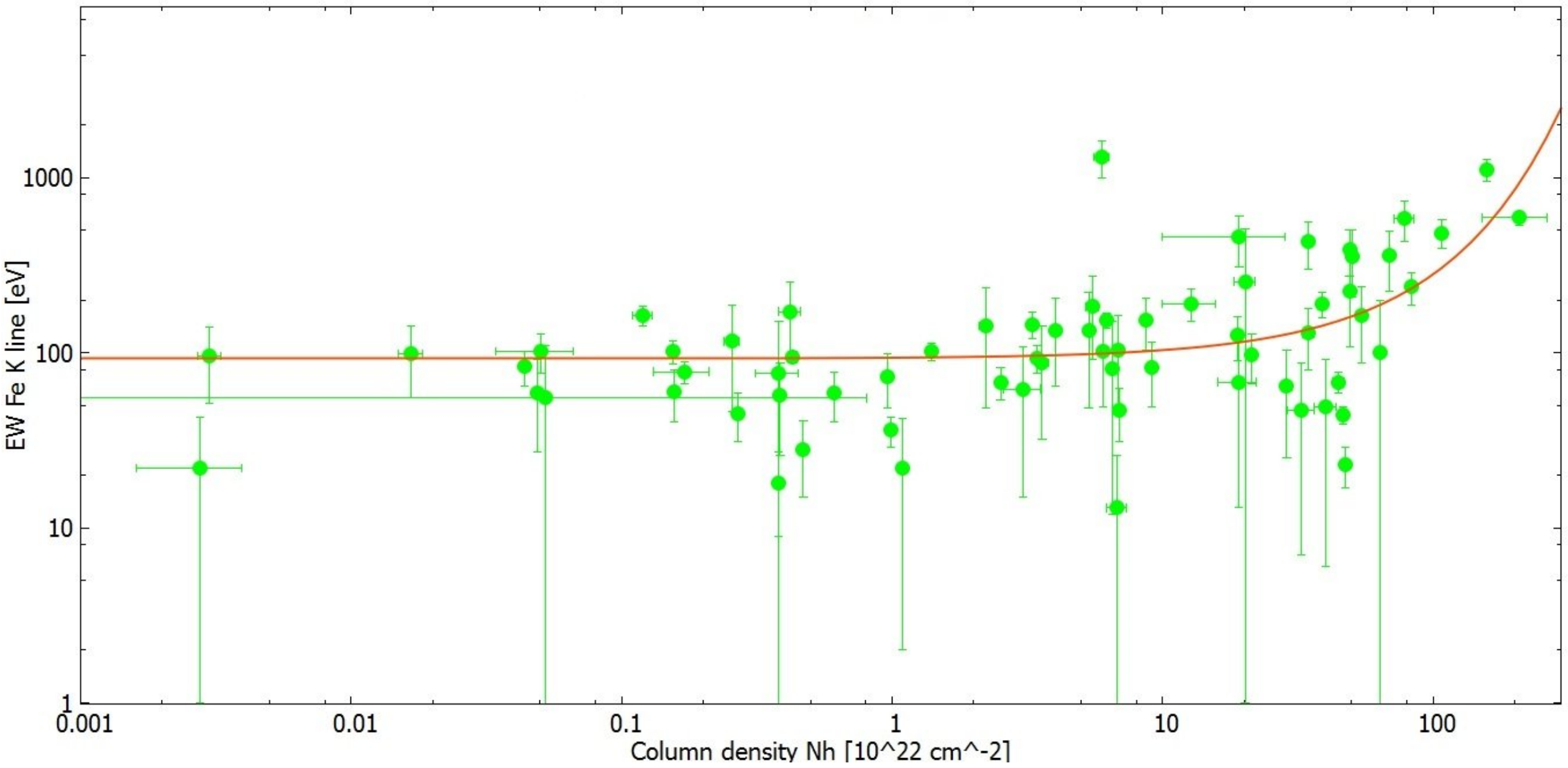

**Fig. 3**. The equivalent width of the Fe $K_\alpha$ line vs the hydrogen column density.

*4.3 Cut-off energy and the photon index*

A possible existence of a correlation between the cutoff energy and the power-law photon index for the first time has been pointed out by Perola et al. (2002); Petrucci et al. (2001); Piro (1999) using BeppoSAX data. But the authors of these works used a very small samples of Seyfert1s – six in (Petrucci et al. 2001) and nine in (Perola et al. 2002) and only four galaxies in (Piro 1999). Later this dependence has been studied using larger amount and better quality data (Dadina 2008; De Rosa et al. 2012; Molina et al. 2008; Molina et al. 2009; Vasudevan et al. 2013, Molina et al. 2013). It is important to understand, either such correlation has a physical origin and or it is an interplay of errors. Indeed, in the data of different authors concerning parameters $\Gamma$ and $E_{cut-off}$ one can see that in many cases, higher values of $\Gamma$ correspond to higher values of $E_{cut-off}$ (see e.g. Perola et al. 2002; Piro 1999). Since the primary spectrum is typically approximated by the formula $E^{-\Gamma}\exp(-E/E_{\text{cut-off}})$, then in a presence of the measurement errors the erroneous increasing of either $\Gamma$ or $E_{cut-off}$ could be compensated (also within error) by an increase of the other parameter. This statistical effect can mask a real physical relationship between these parameters.

We used the data for 57 galaxies, of which 15 are radio-loud. Radio galaxy Mrk 3 was excluded from our analysis because it has an unusual value of the photon index of -0.66, which is quite different from those for the other galaxies. We found the Spearman coefficient is $\rho = 0.31 \pm 0.13$ for the entire sample. For radio-loud galaxies only we obtain $\rho = 0.43 \pm 0.25$; for radio quiet ones $\rho = 0.27 \pm 0.15$. It is widely accepted, that physically reasonable range of the photon index in the X-ray generation scenario by inverse Compton mechanism in the accretion disc coronae (see, e.g., Zdziarski et al. 1990) is between values $1.5 < \Gamma < 2.2$. In this connection, we restricted our sample in conformity with this range of the photon index. So, our sample has been reduced to 39 galaxies and $\rho = 0.37 \pm 0.16$. Thus, our result indicates a lack of a strong correlation and this is consistent with the works of the other authors (see e.g., Molina et al. 2009).

*4.4.The X-ray Baldwin effect*

We also examined the well known “X-ray Baldwin effect” (Baldwin 1977; Iwasawa 1993) which consists in anticorrelation between intrinsic (corrected for the absorption) luminosity $L_{corr}$ and the equivalent width of Fe $K_\alpha$ $EW_{FeK}$ for radio-quiet Seyfert 1s. The existence and explanation of the effect is a subject of discussions. Sometimes it is either non confirmed or found only for certain classes of AGNs (see e.g. Dadina 2008, De Rosa et al. 2012, Winter et al. 2009, Ballantyne, 2014). On the other hand, the effect the decreasing dependence of $EW_{FeK}$ of the luminosity has been claimed (Bianchi et al. 2007; Dadina 2008; Page et al. 2004, Ricci 2013, Gimenez-Garcia et al, 2015) for the Seyfert1s in the range of $2 \div 10$ keV. Recently, Baldwin effect has been reported for Seyfert2s using the Suzaku data (Fukuzawa et al. 2011) and both for Suzaku and XMM-Newton data (Ricci et al. 2014b and references therein).

In our sample of data on 41 galaxies this effect is confirmed, however with low statistical confidence. In the left panel of Fig. 4 we plot the dependence of these parameters for the energy range 2-10 keV. The linear fit shown dependency

$$\log(EW) = (-0.18 \pm 0.05) \times \log(L_{corr}) + (9.81 \pm 2.38)\ .$$

The Spearman correlation coefficient in this case is $\rho = -0.55 \pm 0.12$, that is it indicates a moderate anticorrelation. For 20-100 keV energy (Fig. 4, right panel) relationship has the form

$$\log(EW) = (-0.14 \pm 0.06) \times \log(L_{corr}) + (8.08 \pm 2.54)$$

and Spearman correlation coefficient is $\rho = -0.40 \pm 0.15$.

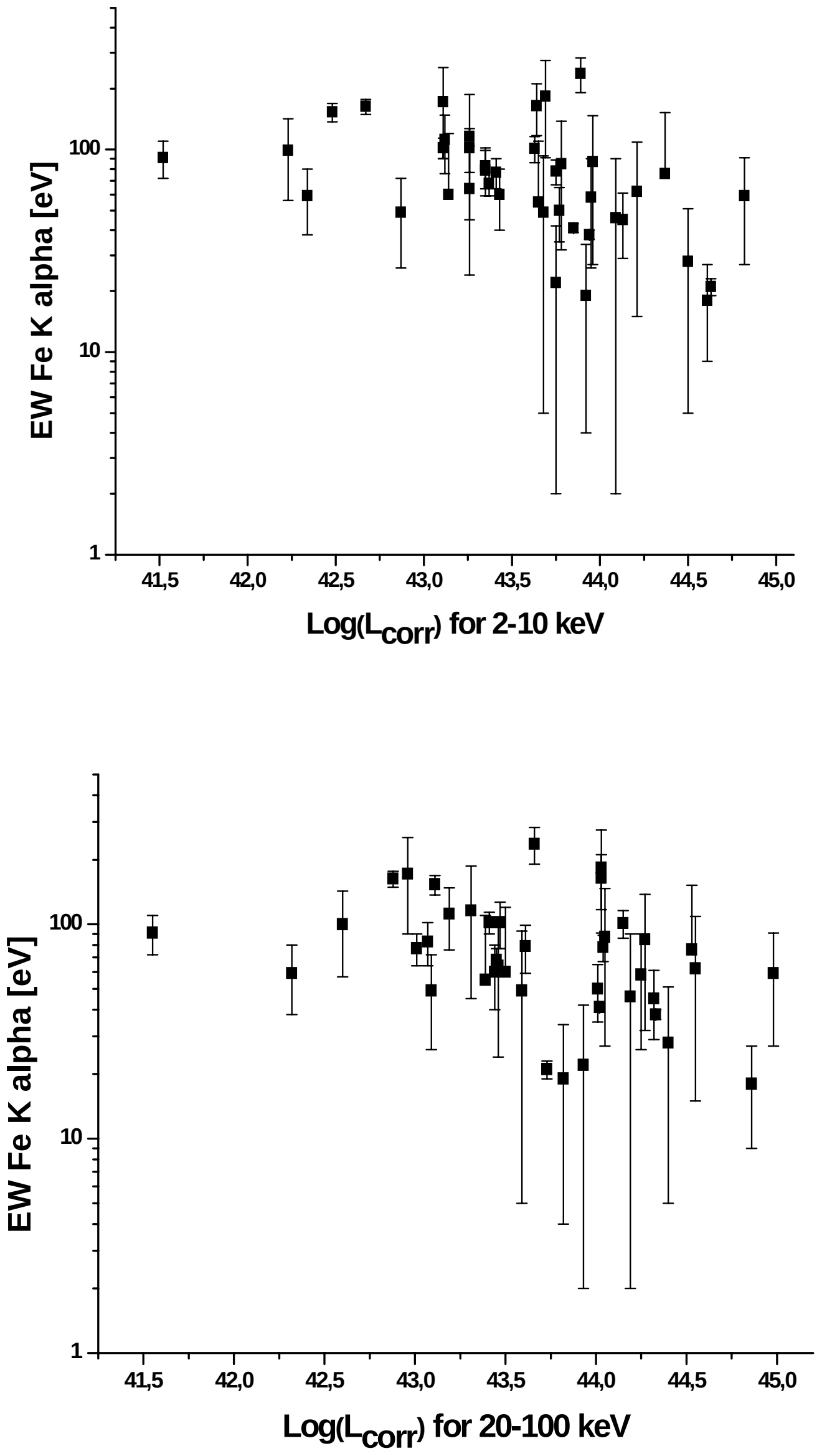

**Fig. 4.** The EW of Fe $K_\alpha$ line as a function of the 2-10 keV (left panel) and 20-100 keV (right panel) luminosity corrected for absorption (the X-ray Baldwin effect).

The physical explanation for this phenomenon is still unclear. Several possibilities are invoked in the literature, such as a change of the ionizing continuum and gas metallicity (in particularly, iron abundance) (Korista et al. 1998) with luminosity or a luminosity-dependent covering factor (Page 2004, Ricci et al. 2013, 2014a and references therein) and ionization state parameter of the BLR (Mushotzky & Ferland 1984) or light bending scenario (for broad component of Fe $K_\alpha$ line) (Miniutti & Fabian 2004).

One of the explanations deals with the covering factor of torus, if we assume that the torus is the source of the iron line emission. Physically, this leads to a change of the torus opening angle, in particularly, to its decreasing and therefore increasing of the sublimation radius and decreasing the thickness of the torus (Grandi et al. 2006; Kawaguchi 2012).

Note that the recent studies often use the accretion rate, rather than the corrected luminosity (see e.g., Winter et al. 2009, 2012).

## 5. X-RAY SPECTRAL PROPERTIES AND UNIFICATION SCHEME

By studying statistics of the main spectral parameters, we can verify their conformity to the UM for AGN (Antonucci 1993). Here, the key parameters are those that describe the continuum and absorption properties. In the AGN unified model the shape of the continuum is expected to be independent of the orientation angle at which the source is observed. Thus, there should be no difference in measured values of the continuum spectral parameters between these types of galaxies. And there should be the only difference in the value of absorption, since for different orientations with respect to the line of sight, the radiation can or cannot pass through the gas-dust torus or its equivalent. The $N_H$ value should be greater for Seyfert 2 types than the average in other AGNs. Differences between the X-ray spectra of various Seyfert types are believed to depend on two factors: 1) the variability of the emission from the corona of the accretion disk; 2) the influence of physical conditions and the type of material that defines the contribution of reflection spectrum, thus affecting the general features of the spectrum. In this section, we analyze the following main spectral parameters: photon index Γ, relative reflection R, exponential energy cut-off, the value of the column densities and equivalent width of the line Fe $K_\alpha$.

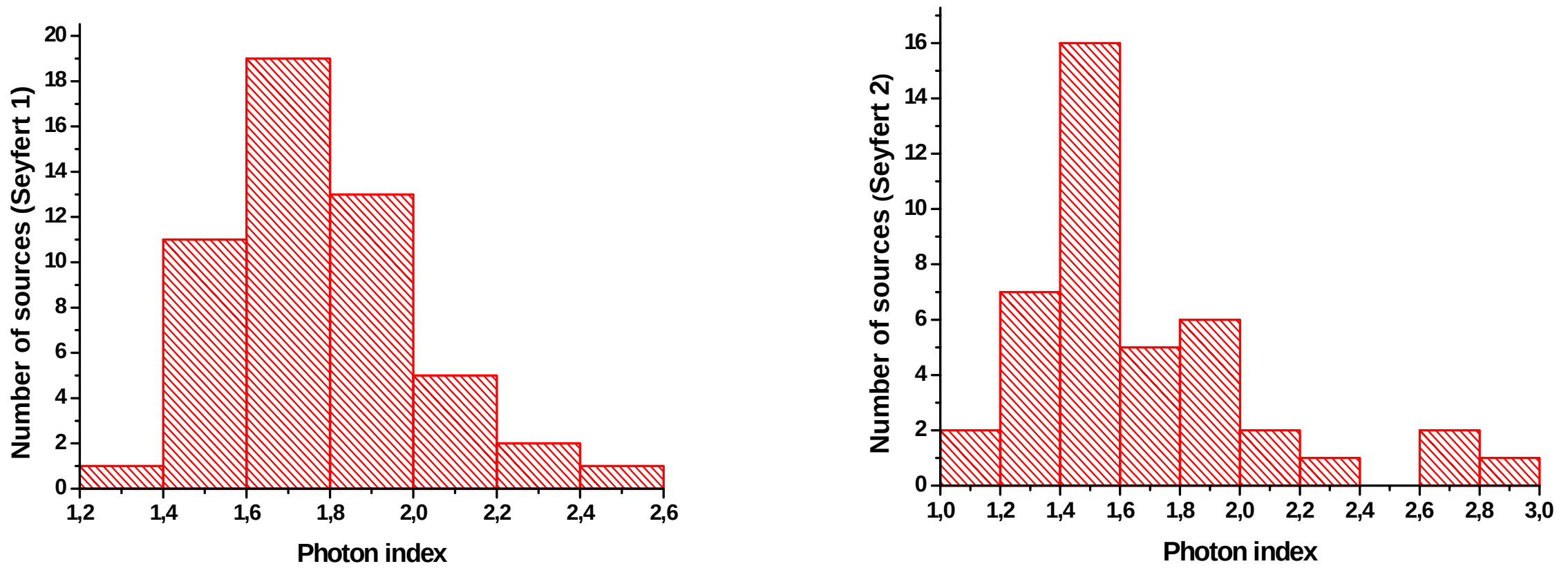

**Fig. 5.** Photon index Γ distribution for the whole dataset.

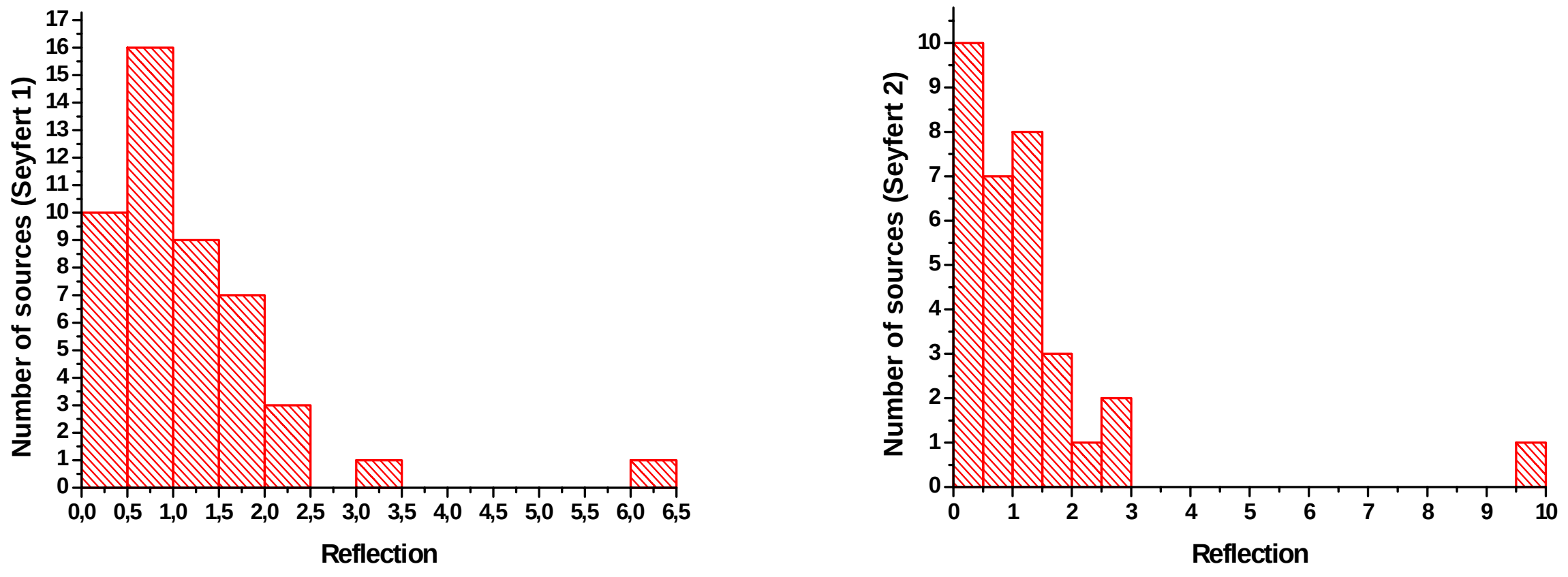

**Fig. 6.** Relative reflection R distribution for the whole dataset.

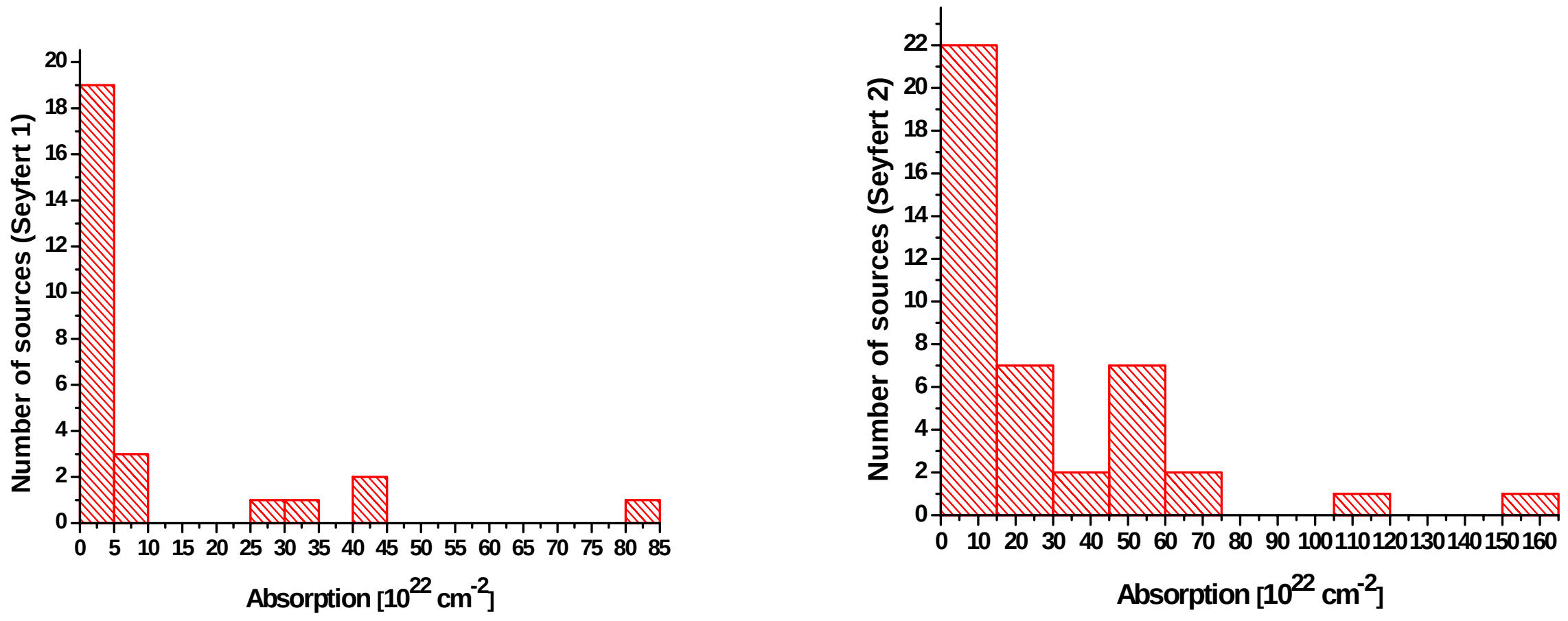

**Fig. 7.** Column density $N_H$ distribution for the whole dataset.

### *5.1 Power-law continuum*

Photon index is the most important parameter of the continuum. As shown in Fig. 5, the Γ value lies in the range of 1.44–2.5 for type 1, mainly concentrating in the range of 1.6-2.0. The average value for type 1 is Γ = 1.78 ± 0.04, which is consistent with other authors (Molina et al. 2009; Panessa et al. 2008; Vasudevan et al. 2013; Winter et al. 2012). For type 2 we obtain wider range of Γ ~ 1.18-2 with average Γ = 1.65 ± 0.05, which is lower than for 1 type and also consistent with other authors [see, e.g., Cappi et al. 2006). Thus, the photon indexes for different types of active galactic nuclei are slightly different, but at the same time, most values lying in the typical range 1.5-2.2. We also used the Kolmogorov-Smirnov test. Accepting the null hypothesis as assuming of absence significant differences between the values of Γ for the different Seyfert types, we got the value of probability $1.2 \cdot 10^{-4}$ (see. Table 1). Therefore, the distributions of photon indexes for various types have significant differences and are statistically different.

The flattening of the photon index (as compared to the primary spectrum) in our sample is mainly observed in Seyfert2s. Discussions in the literature attribute this effect mainly to the fact that existing models used to process X-ray spectra, cannot take into account all complicated spectral features. In particular, the explanation of the flattening for small values of Γ may be the effect of strong absorption and its complexity (see e.g. Panessa et al. 2008; Vasudevan et al. 2013; Winter et al. 2012). In this case the power-law index is distorted possibly due to inadequate accounting the absorption features in the existing models. Also there may be an effect of wrong accounting of a reflection or the presence of of $E_{cut-off}$ in the energy range less than $\sim 100\,\mathrm{keV}$ (see e.g. De Rosa et al. 2008 and refs. therein, Malizia et al. 2009, Georgantopoulos et al. 2013). NGC 1194, NGC 788, NGC 4507 and NGC 4992 of Tables 3 and 4 are typical representatives of this effect. Note that Gondek et al. (1996), Zdziarski et al. (1995) and Zdziarski (2000) found the flattening of the spectra for galaxies with strong absorption, and our data are consistent with these results.

The other possible reason for the flattening is the difference between the accretion disk corona states, especially in the value of Compton parameter (see, e.g., Petrucci et al. 2001). In addition to the influence of absorption and state of corona, the photon index is also believed to be affected by the accretion rate, namely, the increased rate leads to the increased value of Γ (Vasudevan et al. 2013; Winter et al. 2012). Also, for Seyfert 1 spectra (and, to a lesser extent, for Seyfert 2) in explaining Γ<1.6 one must pay attention to the presence of soft excess and the absorption material in different ionization states for energies 0.5-2.0 keV (Panessa et al. 2008; Winter et al. 2012). This is important for fitting the spectral models that cover a range of both medium and soft X-rays, and in our work we do this for the 0.5-2.5 keV range.

*5.2 Relative reflection*

Our spectral results do not give a clear picture of those structures that are mainly responsible for the reflection spectrum in different AGN types. If we assume that the main source of the reflection spectrum is due to the accretion disk, then it must depend only on inclination angle. Then it should be expected that R should be greater in type 1 Seyfert compared with type 2 (Ricci et al. 2011). But in reality this is not observed. For Seyfert 2 with absorption $\geq 10^{23}$ $cm^{-2}$ the value of the relative reflection is bigger. Currently, the main the reason for this influence indicates the reflection from gas-dusty surrounding torus, namely that reflection is bigger for objects with higher inclination angle than for objects with lower one (Awaki et al. 1991; Ricci et al. 2011; Singh et al. 2011; Ueda et al. 2007). As a result, sometimes, on the average, in the 2-10 keV range only pure reflection spectrum from the torus walls can be observed, but not from the accretion disk (Awaki et al. 1991; Singh et al. 2011). If the gas-dusty torus is not a homogeneous structure and consists of individual clouds, we can see a very large value of R (De Rosa et al. 2008; Ricci et al. 2011; Ueda et al. 2007).

Thus, the value of the reflection and its variation cannot be explained only by change of the inclination angle because the core of AGN can contain two main sources of reflection spectrum - the accretion disk and then gas-dusty torus. The first one can show the change of R depending on the ionization state and the state of the accretion flow. The second one depends on the relative motion of possible individual clouds of dust and their size (Beloborodov 1999; Kawaguchi 2012; Magdziarz & Zdziarski 1995; Ueda et al. 2007). In our sample we obtained that the value of reflection for Seyfert 1 galaxy lies in the range 0.05-2.37 (LEDA 090443 and IGR J16558-5203 have unusually high values 6.26 and 3.13, respectively) with a mean value of 1.15 ± 0.15. For Seyfert2s, the variation range is 0.09-2.62 (MCG-03-34-064 has an unusually high value of 9.97) with an average 1.25 ± 0.32, which is higher as compared to type 1, but if we take onto account the errors, then the values of R for both types are similar (see Fig. 6). The large R value for type 2 indicates a possible effect of gas-dust torus, and it is consistent with the results of the earlier studies (Deluit & Courvoisier 2003; Ricci et al. 2011; Singh et al. 2011). Although reflection for Seyfert2s is greater than that for the 1-st type, due to the high errors in the measured values, we cannot claim that this conclusion is final. Kolmogorov-Smirnov two-sample test shows that the probability of the null hypothesis is 0.86 (see Table. 1), i.e. the distributions of R for two Seyfert types are not statistically different.

*5.3 Cut-off energy*

The cutoff energy was first measured for bright Seyfert1s and only last 10 years the data for Seyfert2s has been obtained thanks to sensitive detectors on the satellites. The value of $E_{cut-off}$ is directly related to the temperature and the corona optical depth (Petrucci et al. 2001) that reflect the physical state of the system. The treatment of our samples shows that average $E_{cut-off}$ for 30 galaxies of the 1st type is 154±10 keV, and for 30 galaxies of the 2-d type is 141±10 keV, which is practically the same especially on account of large errors for these values for the individual sources. This invokes doubts concerning a large difference between corona temperatures for different Seyfert types used for explanation of systematically larger $E_{cut-off}$ for Seyfert2s (cf., e.g., Deluit & Courvoisier 2003; Dadina 2008). Moreover, after we ruled out from the sample the radio-loud galaxies that can affect the average value of $E_{cut-off}$, we got $E_{cut-off} = 153 \pm 11\,\mathrm{keV}$ and $152 \pm 9\,\mathrm{keV}$ for Seyfert1s and Seyfert2s respectively. For optical depth $\tau \geq 1$ (Petrucci et al. 2001), this is equivalent to temperature of about $\sim 5.9 \times 10^8\,\mathrm{K}$, which is a typical value. The Kolmogorov-Smirnov test shows the null hypothesis probability to be 0.96 (see Table 1), that is the samples of $E_{cut-off}$ values for different Seyfert types are statistically the same. Therefore, we can state that UM (Antonucci 1993) is valid as regards to the value of $E_{cut-off}$ for different AGN types.

The mean values of the cut-off energy for RL objects are 147 keV for Sy1 type subset and 111.1 keV for Sy2 subset. Note that some objects contradicting to the prediction of the spin-paradigm concerning the cut-off values for RL and RQ AGNs were found. These are RQ AGNs GRS1734-292 (cut-off energy $82^{+11}_{-9}$ keV), IGR J07597-3842 ($71^{+15}_{-11}$), NGC4151 (84 ± 7) and RL M87, 3C 120, IGR J13109-5552, NGC 5128, NGC 5252, NGC1275 (no cut-off), 3C 380.3 ($217^{+78}_{-47}$), 3C 382 ($180^{+63}_{-38}$), 3C 111 ($171^{+29}_{-22}$), 3C 45 ($175^{+134}_{-57}$), NGC 2992 ($232^{+260}_{-88}$). These sources should be recommended for further more detailed investigations of the full set of data available on them to understand exactly the reason of their contradiction with spin paradigm.

### *5.4 Line of sight neutral absorption and equivalent width of Fe $K_\alpha$ line*

The main criterion that determines the Seyfert types in X-rays is the neutral absorption value (see e.g., Ricci et al. 2011; Singh et al. 2011). For our sample this criteria is true; $N_H$ values for type 2 has systematically higher than for type 1 (see Fig. 7); the average value for type 1 is $N_H = 9.62 \pm 7.58 \cdot 10^{22} cm^{-2}$, and for type 2 it is $N_H = 25.86 \pm 9.86 \cdot 10^{22} cm^{-2}$. Kolmogorov-Smirnov test shows that the probability of the null hypothesis $1.7 \cdot 10^{-5}$ (see. Table 1), i.e. the samples of absorption values for different Seyfert types have a significant differences. Therefore, the result of analysis of our sample fully corresponds to UM. However, we note that the average $N_H$ value for type 1 of our sample is rather high because of the presence of the absorbed type 1 AGNs (see, e.g., Malizia et al. 2012), i.e. Seyfert 1s with a high absorbtion. For example, Seyfert 1 galaxy LEDA 75476, which shows a highest absorption for its type, actually is a galaxy with strong variability of column density, which is accompanied by more than 10-fold reduction in the flux (Longinotti et al. 2009). The second by order Sy1 representative with the high absobtion is galaxy NGC 7469. If we remove the absorbed type 1 AGNs from the sample (i.e. with $N_H > 1 \cdot 10^{23} cm^{-2}$), we obtain average $N_H$ of about $1.37 \cdot 10^{22}\, cm^{-2}$. Possible explanations of absorbed Seyferts1s are discussed by, e.g., Malizia et al. (2012) with an emphasis on the existence of an absorbing medium unrelated to the toroidal structure around AGN. We also note that for some galaxies of type 2 there is a shift in the distribution of absorption to smaller values. Some Seyfert 2 galaxies, such as NGC 2992, NGC 2110 and NGC 1365 also show a long term variability in the absorption (Risaliti et al. 2002).

It is important to note that in this view one must have systematically larger $EW_{FeK}$ under the assumption that most narrow Fe lines is formed in the gas-dust torus. The relativistic Fe lines formed in the central parts of the accretion disk also have large $EW_{FeK}$. However, the relativistic effects lead to non-symmetric line profiles that in fact are rarely observed. A consideration of the line generation regions has been carried out in Section 3.2, where a strong relation between $EW_{FeK}$ and $N_H$ has been pointed out. This dependence is not flawed by shifted values of $EW_{FeK}$ which do not correspond (under the reflection assumption) to current value of $N_H$; this justifies and makes reasonable a statistical analysis of the equivalent widths under the assumption that the Fe $K_\alpha$ line is generated in gas-dust tori. Thus for Seyfert1s we found average $EW_{FeK} = 79 \pm 7$ eV within energy band of 13-237 eV. For Seyfert2s we have $EW_{FeK} = 278 \pm 62$ eV within 36-2065 eV. Even if we discard three sources with $EW_{FeK} > 1$ keV (WKK 3050, NGC 4945, NGC 6240), then we have $EW_{FeK} = 180 \pm 24$ eV; this is twice as large than the equivalent width for Seyfert1s. The Kolmogorov-Smirnov test shows the probability of $8.6 \cdot 10^{-4}$ that this is the result of errors (see Table 1). Therefore, the samples for different Seyfert types have reliable statistical differences of absorption. Our results are in accord with the results of Cappi et al. 2006; Singh et al. 2011; Winter et al. 2012; they correspond to the predicted trends of increasing of $EW_{FeK}$ within UM.

## CONCLUSIONS

We have investigated the broad-band X-ray properties of an original sample of 95 AGNs, based on 22-months of hard X-ray Swift/BAT all-sky survey. Namely, we have chosen objects from this survey that have the XMM-Newton data and for which we succeeded to obtain statistically reliable values of spectral parameters using the INTEGRAL data as well. The combined 0.5-250 keV spectra allowed us to perform a study of the emission and absorption features, in particular, to test correlations between spectral parameters that are widely discussed in the literature, such as the reflection coefficient R vs. the photon index $\Gamma$, energy cutoff $E_{cut-off}$ vs. $\Gamma$, hydrogen column density $N_H$ vs. Fe $K_\alpha$ line equivalent width $EW_{FeK}$ and intrinsic luminosity $L_{corr}$ vs. $EW_{FeK}$. The results on correlations between different parameters are given in Table 2. The spectral parameters obtained are presented in Tables 3,4. The Spearman correlation test shows the null hypothesis (that the parameters are completely independent) to be quite unlikely.

The main conclusions of our analysis can be summarized as follows.

1. We found a somewhat higher value reflection parameter R for the photon index values $\Gamma < 1.5$ for Seyfert type 2 as compared to Seyfert type 1 (see Fig.1). This can indicate the influence of an additional reflection from the gas-dust torus. We have not found a strong correlation between photon index $\Gamma$ and relative reflection R; though there exists some increase of $R$ with $\Gamma$. The measured $\Gamma$-R relationship does not fully support the model of the bulk motion of the hot comptonizing plasma, considered by Beloborodov (1999). It means, that the reflected components are not completely due to a production of radiation by reflection from the accretion disc. A possible solution may be that, at least some of the reflection spectra of the Seyfert2s are generated by a dusty tors, which are not continuous, but consist of a clumpy matter, as, e.g., in case of NGC 4945 (Yaqoob 2012), Circinus galaxy (Marinucci et al. 2013) or NGC 3281 (Vasylenko et al. 2013).

2. Though photon index $\Gamma$ and the cut-off energy $E_{cut-off}$ cannot be considered as independent (see Table 2), we have not found a considerable correlation between them. This may be a result of complicated dependence that involves different additional factors as well as due to errors and an insufficient amount of data used.

3. The value of the $EW_{FeK}$ as a function of the absorbing column density $N_H$ is in accord with the conclusion about the effect of the torus obscuring the Fe K line emission. In particular, at least a fraction of the Fe line is to be produced in a Compton-thin environment, likely thin torus or BLR (this still remains the subject of discussions). At the same time, some fraction of this line produced by Compton-thick material, that either can be on the line of sight or not.

4. We found an evidence of the Baldwin effect in Seyfert 1 galaxies, i.e. the decreasing dependence of the equivalent width of the Fe $K_\alpha$ line against the luminosity, in the 2-20 keV band as well as in the 20-100 keV band. However, the statistical significance of this dependence obtained by our data, is not high.

5. We note that our data can be used to investigate the distribution of the main X-ray spectral parameters for different orientations of AGNs so as to test basic assumption of the unified scheme for AGN. The results of this analysis are in a good agreement with the predictions of UM for AGN. No significant differences are obtained while absorption values seem to be the only separator between different types of Seyfert galaxies. In addition to the latter, we found that the equivalent widths of Fe K line of type 2 AGNs have a systematically higher value (~36-2065 eV) than that of type 1 AGNs (~13-237 eV). The values of the high-energy cut-off for Seyfert1s and Seyfert2s have been found to be statistically the same within errors (~153 keV), no matter radio-loud or radio quiet galaxies.

***Acknowledgements***

The authors acknowledge helpful discussions with Dr. Sci. (Hab.) B.I. Hnatyk. We are thankful to the anonymous referee for stimulating comments. This work is partly supported in frame of the Target Program of the Scientific Space Research of the NAS of Ukraine (2012-2016). This research has made use of data and software obtained through the XMM-Newton and INTEGRAL science data centers (ESA); the NASA/IPAC Extragalactic Database operated by the Jet Propulsion Laboratory (CalTech); the HEASARC online service, provided by the NASA/Goddard Space Flight Center. We are thankful to the VIRGO.UA centre in Kiev for offering informational and technical facilities. This work has been supported in part by scientific program "Astronomy and Space Physics", Taras Shevchenko National University of Kyiv.

**Table 1.** Results of Kolmogorov-Smirnov two-sample tests for the statistical comparison of spectral parameters of these two Seyfert types; $D$ is the largest absolute deviation of the two empirical distribution functions, $P$ is the probability that the null hypothesis is correct i.e., that two samples are statistically the same.

| Distribution | Median | | $D$ | $P$ |
|---|---|---|---|---|
| | Seyfert 1s | Seyfert 2s | | |
| Γ | 1.76 | 1.55 | 0.44 | $1.2 \cdot 10^{-4}$ |
| R | 0.97 | 0.9 | 0.12 | 0.86 |
| $E_{cut\text{-}off}$ | 108 | 101 | 0.13 | 0.96 |
| $EW_{FeK}$ | 71 | 129 | 0.42 | $8.6 \cdot 10^{-4}$ |
| $N_H\ (10^{22} cm^{-2})$ | 0.42 | 12.59 | 0.57 | $1.7 \cdot 10^{-5}$ |

**Table 2.** Correlations of parameters: number N of AGN involved, Pearson correlation coefficient P with the jackknifed 1-sigma error, Spearman rank correlation coefficient and probability α of the hypothesis that the parameters are completely independent. The correlations $\Gamma$–$E_{cut\text{-}off}$ are given separately for radio-loud (RL), radio-quiet (RQ), for the restricted file (see the text), and for all 57 AGNs for which both statistically acceptable $\Gamma$ and $E_{cut\text{-}off}$ were available; correlations $\Gamma$–$R$ are given separately for Sy1, Sy2 and total sample of 78 objects; correlations $EW$–$L_{corr}$ are presented (I) for data in the energy range 2-10 keV and (II) for the energy range 20-100 keV.

| | N | P | ρ | α |
|---|---|---|---|---|
| $EW_{FeK}$–$N_H$ | 68 | 0.56±0.24 | 0.46±0.12 | <1% |
| $\Gamma$–$E_c$ | 57 | 0.24±0.14 | 0.31±0.13 | <1% |
| $\Gamma$–$E_{cut\text{-}off}$ (RL) | 15 | 0.44±0.24 | 0.43±0.25 | 5% |
| $\Gamma$–$E_{cut\text{-}off}$ (RQ) | 42 | 0.17± 0.16 | 0.27±0.15 | 5% |
| $\Gamma$–$E_{cut\text{-}off}$ (restr) | 39 | 0.40± 0.16 | 0.37±0.16 | 1% |
| $\Gamma$–$R$ | 78 | 0.53± 0.10 | 0.36±0.11 | <1% |
| $\Gamma$–$R$ (Sy1) | 47 | 0.59±0.11 | 0.47±0.13 | <1% |
| $\Gamma$–$R$ (Sy2) | 31 | 0.60±0.37 | 0.22±0.20 | 12% |
| $EW$–$L_{corr}$ (I) | 41 | -0.36±0.15 | -0.55±0.11 | <1% |
| $EW$–$L_{corr}$ (II) | 41 | -0.29±0.12 | -0.41±0.12 | <1% |

**Table 3.** The best fit spectral parameters – Seyfert type 1

| Name | Γ | R | $E_{cut-off}$, keV | $E_{FeK}$ line, keV | $\sigma_{FeK}$, eV | EW $FeK_{\alpha}$, eV | $L_{corr}^{2-10keV}$, erg/s | $L_{corr}^{20-100keV}$, erg/s | nH, $10^{22}$ $cm^{-2}$ | $\chi^2$ / dof |
|---|---|---|---|---|---|---|---|---|---|---|
| IGR J18027-1455 | 1.52±0.03 | $0.63_{-0.38}^{+0.68}$ | $110_{-38}^{+106}$ | $6.425_{-0.039}^{+0.035}$ | 10 | 116±80 | 1.83e+43 | 2.04e+43 | 0.26±0.02 | 1.008 |
| RX J2135.9+4728 | 1.86±0.02 | 2.43±0.32 | $219_{-111}^{+263}$ | $6.391_{-0.087}^{+0.037}$ | $179_{-79}^{+182}$ | 172±82 | 1.31e+43 | 9.12e+42 | 0.42±0.04 | 0.998 |
| NGC 7314 | 2.05±0.01 | 1.82±0.11 | $246_{-70}^{+155}$ | 6.404±0.042 | $106_{-50}^{+61}$ | 59±19 | 2.17e+42 | 2.08e+42 | 0.61±0.01 | 1.013 |
| NGC 5548 | 1.59±0.01 | 0.89±0.06 | $202_{-34}^{+51}$ | 6.411±0.017 | $74_{-26}^{+30}$ | 77±13 | 2.59e+43 | 1.03e+43 | - | 1.069 |
| WKK 1263 | 1.63±0.03 | $1.51_{-0.60}^{+0.66}$ | $119_{-40}^{+104}$ | $6.562_{-0.468}^{+0.064}$ | 10 | $60_{-60}^{+73}$ | 1.41e+43 | 3.18e+43 | - | 1.011 |
| IGR J16558-5203 | 2.31±0.02 | 3.13±0.47 | - | 6.41 | 10 | $47_{-47}^{+43}$ | 1.25e+44 | 1.56e+44 | $32.67_{-3.76}^{+4.41}$ | 1.097 |
| GRS 1734-292 | 1.52±0.01 | $0.38_{-0.19}^{+0.15}$ | $82_{-9}^{+11}$ | 6.394±0.538 | 10 | $22_{-22}^{+20}$ | 5.65e+43 | 8.52e+43 | 1.09±0.02 | 0.962 |
| Mrk 110 | 1.59±0.01 | 0.24±0.08 | - | $6.425_{-0.027}^{+0.033}$ | 10 | 38±1 | 8.77e+43 | 2.14e+44 | - | 1.057 |
| Mrk 926 | 2.01±0.02 | $1.95_{-0.88}^{+0.95}$ | - | 6.387±0.097 | $118_{-75}^{+109}$ | 62±47 | 1.63e+44 | 3.52e+44 | $3.05_{-0.48}^{+0.62}$ | 0.979 |
| MCG+8-11-11 | 1.79±0.01 | 1.86±0.23 | - | 6.426±0.019 | 36±35 | 101±16 | 4.25e+43 | 1.42e+44 | 0.15±0.02 | 0.946 |
| NGC 7469 | 2.01±0.01 | 0.23±0.02 | - | 6.425±0.016 | $39_{-39}^{+23}$ | 68±9 | 2.35e+43 | 2.81e+43 | $44.78_{-2.29}^{+2.46}$ | 1.098 |
| NGC 3783 | 1.69±0.01 | 0.81±0.17 | $167_{-35}^{+58}$ | 6.411±0.007 | $54_{-12}^{+8}$ | 102±12 | 1.30e+43 | 2.58e+43 | - | 1.377 |
| 1A 1143-18 | 1.67±0.01 | 0.97±0.24 | $128_{-24}^{+37}$ | 6.426±0.046 | $55_{-47}^{+59}$ | 41±1 | 7.17e+43 | 1.06e+44 | - | 0.951 |
| IGR J16482-3036 | $1.59_{-0.02}^{+0.04}$ | $1.49_{-0.68}^{+1.09}$ | $101_{-37}^{+150}$ | $6.438_{-0.077}^{+0.101}$ | 10 | $55_{-55}^{+73}$ | 4.47e+43 | 2.47e+43 | - | 1.089 |
| IGR J17488-3253 | 1.44±0.02 | 0.18±0.18 | $135_{-49}^{+143}$ | - | - | - | 1.39e+43 | 2.89e+43 | 0.07±0.02 | 0.939 |
| SWIFT J1038.8-4942 | 1.51±0.04 | 1.25±0.38 | $102_{-41}^{+128}$ | 6.407±0.036 | 10 | $183_{-99}^{+92}$ | 4.97e+43 | 1.09e+44 | $5.50_{-0.30}^{+0.33}$ | 1.140 |
| LEDA 090443 | 2.28±0.04 | $6.26_{-1.17}^{+1.36}$ | $134_{-44}^{+122}$ | 6.446±0.054 | $63_{-63}^{+93}$ | $87_{-60}^{+56}$ | 9.14e+43 | 1.14e+44 | 3.59±0.19 | 0.971 |
| IGR J07597-3842 | 1.56±0.01 | $0.61_{-0.33}^{+0.35}$ | $71_{-11}^{+15}$ | 6.430±0.070 | $119_{-69}^{+87}$ | $85_{-58}^{+53}$ | 6.04e+43 | 1.86e+44 | - | 0.924 |
| LEDA 168563 | 1.66±0.01 | 0.11±0.07 | $159_{-40}^{+78}$ | 6.479±0.107 | 10 | 19±16 | 8.42e+43 | 6.62e+43 | - | 1.009 |
| NGC 4593 | 1.76±0.01 | 0.69±0.11 | $122_{-38}^{+68}$ | 6.402±0.012 | 78±18 | 119±18 | 7.49e+42 | 1.26e+43 | - | 1.043 |
| NGC 7603 | 2.11±0.01 | 0.99±0.38 | - | $6.423_{-0.075}^{+0.054}$ | 10 | $49_{-44}^{+49}$ | 4.87e+43 | 3.91e+43 | $40.09_{-5.47}^{+6.68}$ | 1.093 |
| ESO 511-30 | 1.83±0.01 | 1.31±0.15 | - | 6.404±0.024 | $95_{-36}^{+48}$ | 80±20 | 2.27e+43 | 4.13e+43 | - | 1.058 |
| NGC 985 | 1.49±0.01 | 1.04±0.26 | - | 6.438±0.031 | $113_{-33}^{+46}$ | $164_{-51}^{+47}$ | 4.40e+43 | 1.08e+44 | - | 1.061 |

**Table 3.** (Continued)

| Name | $\Gamma$ | R | $E_{cut-off}$, keV | $E_{FeK}$ line, keV | $\sigma_{FeK}$, eV | EW $FeK_{\alpha}$, eV | $L_{corr}^{2-10keV}$, erg/s | $L_{corr}^{20-100keV}$, erg/s | nH, $10^{22}$ $cm^{-2}$ | $\chi^2$ / dof |
|---|---|---|---|---|---|---|---|---|---|---|
| ESO 140-43 | 1.97±0.01 | $0.66^{+0.28}_{-0.22}$ | - | $6.399^{+0.033}_{-0.030}$ | $87^{+39}_{-35}$ | $112^{+37}_{-42}$ | 1.34e+43 | 1.55e+43 | - | 1.190 |
| 2E 1853.7+1534 | 1.89±0.03 | $2.24^{+0.93}_{-0.81}$ | $129^{+232}_{-56}$ | $6.365^{+0.064}_{-0.081}$ | 10 | $76^{+79}_{-76}$ | 2.36e+44 | 3.43e+44 | 0.38±0.07 | 0.965 |
| 1A 1343-60 | 1.65±0.02 | 0.58±0.19 | $110^{+37}_{-21}$ | 6.412±0.032 | 26±26 | 64±40 | 1.83e+43 | 2.89e+43 | $28.87^{+22.94}_{-12.41}$ | 1.036 |
| 4U0517+17 | 1.74±0.01 | $1.91^{+0.28}_{-0.23}$ | - | 6.415±0.021 | 77±36 | $102^{+25}_{-27}$ | 1.84e+43 | 2.96e+43 | - | 0.977 |
| NGC 931 | 1.78±0.01 | 1.08±0.38 | $206^{+141}_{-105}$ | 6.434±0.019 | $86^{+51}_{-22}$ | 83±20 | 2.27e+43 | 1.19e+43 | - | 1.155 |
| NGC 6814 | 1.72±0.01 | $0.51^{+0.27}_{-0.25}$ | - | 6.404±0.021 | 47±31 | $99^{+50}_{-33}$ | 1.72e+42 | 4.41e+42 | 0.02±0.01 | 1.014 |
| IC4329A[1] | 1.73±0.01 | $0.57^{+0.06}_{-0.10}$ | $196^{+42}_{-30}$ | 6.395±0.011 | $81^{+15}_{-13}$ | 78±11 | 5.63e+43 | 1.10e+44 | 0.17±0.04 | 1.072 |
| MR 2251-178 | 1.63±0.01 | 0.34±0.09 | $206^{+54}_{-36}$ | $6.431^{+0.030}_{-0.036}$ | 10 | $18^{+11}_{-9}$ | 4.03e+44 | 7.36e+44 | 0.38±0.01 | 1.038 |
| ESO 141-55 | 2.01±0.01 | 0.74±0.15 | - | 6.410±0.029 | 10 | 57±32 | 8.85e+43 | 1.76e+44 | 0.38±0.01 | 1.147 |
| ESO 490-26 | 1.64±0.01 | - | $158^{+124}_{-51}$ | 6.424±0.058 | 10 | $60^{+20}_{-23}$ | 2.76e+43 | 2.76e+43 | 0.15±0.01 | 0.949 |
| PGC 045125 | 1.84±0.01 | - | - | $6.437^{+0.225}_{-0.204}$ | 10 | 22±1 | 4.11e+43 | 5.45e+43 | - | 0.912 |
| Ark 120 | 1.86±0.01 | 1.69±0.08 | - | 6.442±0.016 | 10 | 50±16 | 5.82e+43 | 1.01e+44 | - | 1.132 |
| NGC 4051 | 1.93±0.01 | 0.92±0.09 | $237^{+115}_{-59}$ | 6.404±0.014 | $47^{+28}_{-46}$ | $91^{+19}_{-21}$ | 3.30e+41 | 3.54e+41 | - | 0.988 |
| Mrk 509 | 1.97±0.01 | 0.39±0.06 | - | 6.429±0.020 | 10 | 45±14 | 1.34e+44 | 2.11e+44 | 0.27±0.01 | 1.079 |
| MCG -6-30-15 | 1.89±0.01 | $1.18^{+110}_{-0.07}$ | $247^{+74}_{-47}$ | 6.421±0.104 | $704^{+151}_{-119}$ | $163^{+14}_{-38}$ | 4.69e+42 | 7.55e+42 | 0.12±0.01 | 1.112 |
| NGC 4151 | 1.28±0.02 | 0.05±0.03 | 84±7 | 6.435±0.007 | 48±13 | 153±16 | 3.05e+42 | 1.30e+43 | 6.23±0.07 | 1.073 |
| LEDA 75476 | 1.97±0.03 | - | - | 6.432±0.012 | $20^{+28}_{-20}$ | 237±50 | 7.81e+43 | 4.55e+43 | $83.75^{+5.13}_{-4.93}$ | 0.973 |
| M 87* | 2.50±0.01 | - | - | - | - | - | 8.81e+41 | 3.88e+40 | - | 1.117 |
| 4C 74.26* | 1.76±0.01 | 1.24±0.19 | 76±17 | 6.416±0.074 | $167^{+74}_{-52}$ | 59±33 | 6.64e+44 | 9.74e+44 | 0.05±0.02 | 1.041 |
| IGR J13109-5552* | 1.58±0.05 | $1.58^{+1.35}_{-1.15}$ | - | 6.436±0.529 | 10 | $96^{+153}_{-96}$ | 9.33e+43 | 4.16e+44 | - | 0.899 |
| 3C 382* | 1.76±0.01 | 0.68±0.19 | $180^{+63}_{-38}$ | 6.397±0.044 | 10 | 28±24 | 3.17e+44 | 2.53e+44 | - | 0.994 |
| 4C 50.55* | 1.45±0.01 | 0.54±0.13 | $91^{+11}_{-9}$ | $6.444^{+0.069}_{-0.079}$ | 10 | $13^{+18}_{-13}$ | 5.25e+43 | 1.43e+44 | $6.76^{+0.99}_{-0.87}$ | 1.013 |
| 3C 111* | 1.60±0.01 | 0.29±0.11 | $171^{+29}_{-22}$ | 6.419±0.029 | 48±33 | 28±13 | 2.76e+44 | 3.77e+44 | 0.47±0.01 | 1.053 |
| 3C 390.3* | 1.78±0.01 | 1.28±0.17 | $217^{+78}_{-47}$ | 6.421±0.053 | 10 | 25±19 | 2.96e+44 | 3.67e+44 | - | 1.026 |

**Table 3.** (Continued)

| Name | Γ | R | $E_{cut\text{-}off}$, keV | $E_{FeK}$ line, keV | $\sigma_{FeK}$, eV | EW $FeK_{\alpha}$, eV | $L_{corr}^{2-10keV}$, erg/s | $L_{corr}^{20-100keV}$, erg/s | nH, $10^{22}$ $cm^{-2}$ | $\chi^2$ / dof |
|---|---|---|---|---|---|---|---|---|---|---|
| Pictor A* | 1.80±0.01 | 0.99±0.23 | - | $6.423^{+0.504}_{-0.367}$ | 10 | $21^{+39}_{-21}$ | 2.96e+43 | 9.41e+43 | 0.03±0.01 | 0.903 |
| 3C 120* | 1.74±0.01 | $0.25^{+17.79}_{-0.09}$ | - | 6.419±0.017 | 102±23 | 74±13 | 1.16e+44 | 1.46e+44 | - | 1.114 |
| QSO B0241+62* | 1.61±0.01 | 1.05±0.13 | $216^{+161}_{-67}$ | 6.411±0.021 | 83±28 | 126±31 | 9.11e+43 | 2.18e+44 | - | 0.929 |
| S5 2116+81* | 1.96±0.04 | $2.37^{+1.15}_{-0.99}$ | - | - | - | - | 1.82e+44 | 5.27e+44 | - | 0.974 |
| WKK 6471* | 1.91±0.05 | - | - | - | - | - | 4.64e+42 | 3.49e+43 | - | 1.045 |

[1]Analyzed in the 2.5-250 keV range
*Sources classified as Radio loud objects

**Table 4.** The best fit spectral parameters – Seyfert type 2

| Name | Γ | R | $E_{cut\text{-}off}$, keV | $E_{FeK}$ line, keV | $\sigma_{FeK}$, eV | EW $FeK_{\alpha}$, eV | $L_{corr}^{2-10keV}$, erg/s | $L_{corr}^{20-100keV}$, erg/s | nH, $10^{22}$ $cm^{-2}$ | $\chi^2$ / dof |
|---|---|---|---|---|---|---|---|---|---|---|
| MCG-01-24-012 | 1.77±0.04 | $1.88^{+1.22}_{-1.02}$ | $107^{+110}_{-47}$ | 6.400±0.052 | $49^{+92}_{-2}$ | $81^{+74}_{-68}$ | 1.74e+43 | 2.81e+43 | 6.51±0.17 | 0.852 |
| NGC 4074 | $1.82^{+0.43}_{-0.35}$ | - | - | 6.41 | 10 | $255^{+372}_{-254}$ | 7.91e+42 | 7.82e+42 | $20.23^{+2.01}_{-1.69}$ | 1.358 |
| ESO 103-35 | 1.94±0.03 | $1.89^{+0.82}_{-0.69}$ | $119^{+40}_{-26}$ | $6.433^{+0.055}_{-0.087}$ | $66^{+292}_{-66}$ | $68^{+65}_{-46}$ | 2.25e+43 | 2.99e+43 | 19.13±0.31 | 1.111 |
| PGC 037894 | $1.54^{+0.06}_{-0.04}$ | $1.42^{+1.53}_{-1.04}$ | $136^{+231}_{-56}$ | - | - | - | 2.31e+43 | 6.36e+43 | 7.54±0.26 | 0.980 |
| WKK 0560 | $1.22^{+0.10}_{-0.07}$ | $1.44^{+2.97}_{-1.73}$ | $104^{+97}_{-42}$ | $6.378^{+0.066}_{-0.072}$ | $39^{+22}_{-39}$ | $142^{+94}_{-97}$ | 1.42e+43 | 5.47e+43 | 2.23±0.14 | 0.905 |
| NGC 4138 | 1.36±0.03 | $0.09^{+0.72}_{-0.01}$ | $156^{+501}_{-84}$ | 6.41 | $51^{+80}_{-51}$ | $104^{+58}_{-65}$ | 1.50e+41 | 4.66e+41 | $6.85^{+0.27}_{-0.25}$ | 1.003 |
| NGC 1142 | $1.67^{+0.09}_{-0.06}$ | $1.69^{+2.01}_{-1.03}$ | - | $6.421^{+0.028}_{-0.032}$ | $14^{+58}_{-7}$ | $225^{+116}_{-121}$ | 2.57e+43 | 1.11e+44 | $49.49^{+1.72}_{-1.61}$ | 0.917 |
| NGC 1194[1] | 1.18±0.15 | 1.15±0.89 | $99^{+207}_{-35}$ | 6.376±0.041<br>6.514±0.047 | - | $561^{+248}_{-85}$<br>$459^{+86}_{-71}$ | 4.23e+41 | 97.81e+41 | 108.07±4.93 | 0.991 |
| NGC 3281 | 2.04±0.01 | - | - | 6.398±0.070 | $50^{+17}_{-49}$ | 589±54 | 1.84e+42 | 1.27e+43 | 208±52 | 1.084 |
| ESO 506-27 | 1.46±0.01 | 0.73±0.34 | 155±62 | $6.417^{+0.021}_{-0.031}$ | $42^{+38}_{-42}$ | 359±136 | 3.26e+43 | 1.19e+44 | $69.07^{+2.06}_{-1.94}$ | 1.122 |
| IGR J20187+4041 | 1.57±0.04 | $1.48^{+1.31}_{-1.07}$ | $77^{+43}_{-21}$ | 6.339±0.049 | $55^{+73}_{-55}$ | $134^{+86}_{-88}$ | 3.60e+42 | 7.63e+42 | 5.34±0.20 | 0.905 |
| IC 4518A | 1.55±0.01 | 2.62±0.61 | $127^{+663}_{-66}$ | 6.401±0.024 | $79^{+23}_{-24}$ | 456±148 | 1.88e+42 | 9.25e+42 | $19.24^{+0.99}_{-0.92}$ | 0.914 |

**Table 4.** (Continued)

| Name | Γ | R | $E_{cut-off}$, keV | $E_{FeK}$ line, keV | $\sigma_{FeK}$, eV | EW $FeK_{\alpha}$, eV | $L_{corr}^{2-10keV}$, erg/s | $L_{corr}^{20-100keV}$, erg/s | nH, $10^{22}$ $cm^{-2}$ | $\chi^2$ / dof |
|---|---|---|---|---|---|---|---|---|---|---|
| NGC 4945 | 1.80±0.06 | 0.12±0.09 | - | 6.399±0.007 | 10 | $1108^{+163}_{-157}$ | 2.49e+41 | 1.72e+42 | $158.55^{+2.86}_{-2.73}$ | 1.357 |
| NGC 6300 | 1.62±0.02 | $1.00^{+0.42}_{-0.44}$ | - | 6.421±0.016 | $27^{+17}_{-27}$ | $125^{+33}_{-36}$ | 5.96e+41 | 2.21e+42 | 18.95±0.24 | 1.019 |
| NGC 526 | 1.41±0.01 | $0.21^{+0.33}_{-0.31}$ | $214^{+143}_{-63}$ | $6.393^{+0.034}_{-0.022}$ | $67^{+134}_{-44}$ | 73±26 | 1.95e+43 | 4.46e+43 | 0.96±0.01 | 1.089 |
| LEDA 178130 | 1.55±0.03 | $2.09^{+0.79}_{-0.84}$ | $217^{+278}_{-83}$ | 6.388±0.041 | $55^{+49}_{-55}$ | 101±52 | 3.13e+43 | 1.29e+44 | 6.02±0.13 | 0.940 |
| ESO 362-18 | 2.80±0.02 | - | $173^{+128}_{-99}$ | 6.410±0.018 | $95^{+25}_{-21}$ | 190±40 | 3.68e+42 | 1.44e+43 | $12.82^{+4.11}_{-2.84}$ | 1.021 |
| MCG +04-48-002 | $1.46^{+0.16}_{-0.12}$ | - | - | $6.479^{+0.181}_{-0.566}$ | 10 | $100^{+144}_{-100}$ | 7.06e+42 | 2.33e+43 | $63.96^{+2.76}_{-2.53}$ | 0.794 |
| NGC 4992 | 1.41±0.02 | - | $179^{+179}_{-100}$ | 6.386±0.034 | $79^{+44}_{-39}$ | $356^{+149}_{-157}$ | 1.24e+43 | 5.78e+43 | $50.39^{+1.86}_{-1.74}$ | 0.771 |
| NGC 4507 | 1.32±0.02 | $1.04^{+2.19}_{-1.04}$ | $167^{+81}_{-45}$ | 6.381±0.011 | $49^{+19}_{-27}$ | 190±31 | 1.29e+43 | 4.12e+43 | 38.83±0.40 | 1.195 |
| MCG -05-23-016[2] | 1.71±0.01 | 0.42±0.06 | $185^{+27}_{-37}$ | 6.429±0.046 | $318^{+52}_{-45}$ | 36±7 | 2.11e+43 | 2.18e+43 | 0.99±0.03 | 1.036 |
| NGC 5506 | 1.89±0.01 | 0.43±0.05 | $201^{+46}_{-33}$ | 6.405±0.019 | 10 | 93±17 | 1.06e+43 | 1.32e+43 | 3.45±0.01 | 1.043 |
| NGC 4388 | 1.56±0.03 | 0.22±0.10 | $159^{+33}_{-25}$ | 6.393±0.023 | $67^{+32}_{-46}$ | 97±31 | 1.63e+43 | 3.39e+43 | 21.44±0.24 | 1.052 |
| LEDA 93974 | 1.99±0.04 | 2.61±0.48 | - | $6.376^{+0.051}_{-0.046}$ | $99^{+122}_{-49}$ | 134±70 | 2.14e+43 | 5.25e+43 | 4.02±0.04 | 1.048 |
| NGC 6240 | $2.70^{+0.11}_{-0.16}$ | - | $100^{+101}_{-61}$ | $6.396^{+0.025}_{-0.020}$ | $79^{+29}_{-36}$ | $1312^{+327}_{-304}$ | 7.62e+44 | 4.45e+43 | 5.96±0.44 | 1.244 |
| MCG -03-34-064[3] | 2.36±0.09 | $9.97^{+3.30}_{-2.71}$ | $218^{+218}_{-104}$ | 6.412±0.021 (6.434±0.055 ) | $12^{+52}_{-7}$ | 129±49 (431±132) | 5.45e+42 | 9.28e+42 | 34.62±1.63 | 1.065 |
| NGC 7172 | 1.50±0.02 | 0.33±0.17 | - | 6.434±0.023 | 48±42 | 92±32 | 5.43e+42 | 1.98e+43 | 7.96±0.09 | 1.012 |
| NGC 1365 | 2.17±0.02 | 0.77±0.65 | - | 6.388±0.013 | 10 | 78±21 | 2.10e+42 | 2.49e+42 | 12.36±0.38 | 1.287 |
| NGC 3227 | 1.38±0.01 | 0.90±0.15 | $155^{+21}_{-17}$ | 6.416±0.009 | 53±13 | 94±7 | 1.22e+42 | 4.28e+42 | 0.43±0.01 | 1.133 |
| NGC 7582 | 1.77±0.02 | 0.37±0.21 | - | 6.407±0.015 | $34^{+28}_{-34}$ | 581±149 | 4.55e+41 | 2.10e+42 | $15.71^{+1.48}_{-1.34}$ | 1.050 |
| NGC 1052* | 1.41±0.04 | $0.10^{+0.34}_{-0.05}$ | $150^{+202}_{-110}$ | 6.398±0.023 | $72^{+34}_{-50}$ | 154±50 | 4.71e+41 | 9.57e+41 | 8.66±0.17 | 1.126 |
| Mrk 6* | 1.48±0.01 | 0.52±0.13 | $226^{+186}_{-175}$ | 6.449±0.025 | 18±12 | 77±30 | 1.31e+43 | 4.59e+43 | - | 0.993 |
| 3C 405[2]* | 1.51±0.02 | - | $175^{+134}_{-57}$ | - | - | - | 2.24e+44 | 5.64e+44 | 25.01±1.24 | 1.099 |
| NGC 1275* | 2.67±0.04 | 0.59±0.39 | - | - | - | - | 1.52e+43 | 2.38e+43 | 0.85±0.01 | 1.077 |

**Table 4.** (Continued)

| Name | Γ | R | $E_{cut\text{-}off}$, keV | $E_{FeK}$ line, keV | $\sigma_{FeK}$, eV | EW $FeK_{\alpha}$, eV | $L_{corr}^{2-10keV}$, erg/s | $L_{corr}^{20-100keV}$, erg/s | nH, $10^{22}$ $cm^{-2}$ | $\chi^2$ / dof |
|---|---|---|---|---|---|---|---|---|---|---|
| NGC 5128* | 1.82±0.02 | - | - | 6.415±0.019 | 10 | 82±33 | 2.42e+42 | 6.32e+42 | 9.14±0.23 | 1.445 |
| Mrk 3* | $0.66^{+0.03}_{-0.07}$ | - | $63^{+9}_{-7}$ | 6.398±0.007 | 10 | 44±5 | 5.89e+42 | 3.84e+43 | 46.49±1.03 | 1.470 |
| WKK 3050[2]* | 1.18±0.02 | 0.86±0.14 | 42±2 | 6.402±0.002 | 23±6 | 2065±7 | 1.21e+41 | 9.07e+41 | 47.44±0.82 | 1.209 |
| Mrk 348* | 1.55±0.02 | $0.67^{+0.37}_{-0.34}$ | $67^{+9}_{-7}$ | 6.401±0.035 | 10 | 47±16 | 2.81e+43 | 5.23e+43 | 12.21±0.39 | 0.966 |
| 3C 452* | 1.35±0.03 | $1.04^{+0.07}_{-0.14}$ | $37^{+13}_{-8}$ | 6.417±0.025 | 58±30 | $163^{+84}_{-77}$ | 7.99e+43 | 4.76e+44 | $54.52^{+3.12}_{-2.85}$ | 1.123 |
| NGC 788* | $1.23^{+0.06}_{-0.04}$ | $1.08^{+1.21}_{-0.74}$ | $117^{+46}_{-30}$ | $6.440^{+0.019}_{-0.024}$ | 67±29 | $387^{+116}_{-112}$ | 4.75e+42 | 2.71e+43 | $49.10^{+1.70}_{-1.61}$ | 1.089 |
| NGC 2110* | 1.56±0.03 | $0.91^{+0.36}_{-0.29}$ | $101^{+70}_{-33}$ | 6.423±0.011 | $45^{+20}_{-26}$ | 144±26 | 3.84e+42 | 2.72e+43 | 3.32±0.07 | 1.129 |
| NGC 5252* | 1.31±0.01 | - | - | 6.409±0.029 | 37±33 | 69±15 | 1.82e+43 | 1.33e+44 | 2.52±0.04 | 0.984 |
| NGC 2992* | 1.52±0.02 | 0.20±0.18 | $232^{+260}_{-88}$ | 6.431±0.024 | $62^{+36}_{-44}$ | 208±69 | 1.86e+42 | 1.05e+43 | 0.79±0.03 | 0.867 |

[1]Parameters for two relativistic Fe $K_{\alpha}$ line represented by the diskline model: see Vasylenko et al. 2015
[2] Analyzed in the 2.5-250 keV range
[3]Parameters are presented for narrow and broad components of Fe $K_{\alpha}$ line
*Sources classified as Radio loud objects